\documentclass[11pt]{article}

\usepackage{graphicx}
\usepackage{endfloat}
\usepackage{amssymb}

\usepackage{amsmath}

\usepackage[default]{jasa_harvard}    
\usepackage{JASA_manu}

\begin{document}

\title{Computation of ancestry scores with mixed families and unrelated individuals}
\author{Yi-Hui Zhou$^*$  \\ J. S. Marron$^\dagger$ \\ Fred A. Wright$^\ddagger$\\ \\
$^*\ddagger$Bioinformatics Research Center and
Departments of Biological Sciences and $^\ddagger$Statistics, \\
North Carolina State University
  \\ 
$^\dagger$Department of Statistics and Operations Research,\\ University of North Carolina\\
email: \texttt{yzhou19@ncsu.edu} }

\maketitle

\newpage
\begin{center}
\textbf{Abstract}
\end{center}
The issue of robustness to family relationships in computing genotype ancestry scores such as eigenvector projections has received increased attention in genetic association, as the scores are widely used to control spurious association.  We use a motivational example from the North American Cystic Fibrosis (CF) Consortium genetic association study with 3444 individuals and 898 family members to illustrate the challenge of computing ancestry scores when sets of both unrelated individuals and closely-related family members are included.  We propose novel methods to obtain ancestry scores and demonstrate that the proposed methods outperform existing methods. The current standard is to compute loadings (left singular vectors) using unrelated individuals and to compute projected scores for remaining family members. However, projected ancestry scores from this approach suffer from shrinkage toward zero. We consider in turn alternate strategies: (i) within-family data orthogonalization, (ii) matrix substitution based on decomposition of a target family-orthogonalized covariance matrix, (iii) covariance-preserving whitening, retaining covariances between unrelated pairs while orthogonalizing family members, and (iv) using family-averaged data to obtain loadings. Except for within-family orthogonalization, our proposed approaches offer similar performance and are superior to the standard approaches. We illustrate the performance via simulation and analysis of the CF dataset. 
\vspace*{.3in}


\newpage

\section{Introduction}\label{introduction}
Differing ancestries of human subpopulations create systematic differences in genetic allele frequencies across the genome, a phenomenon known as population stratification or substructure.  If a phenotypic trait such as disease is associated with subpopulation membership, a genetic association study can identify spurious relationships with genetic markers. Singular value decomposition (SVD) of genotype data or eigen decomposition of covariance matrices 
can be used to identify population stratification. The eigenvectors (essentially principal component scores) that
correspond to large eigenvalues can be used as covariates in association analysis  \cite{levine2013genome}. 
The combined analysis of unrelated and related individuals is a common feature of genetic association studies
\cite{zhu2008unified}. However, the presence of close-degree relatives in a genetic dataset presents difficulties, as the family structure can greatly influence the eigenvalues and eigenvectors.

\begin{figure}
\begin{center}
\includegraphics[scale=0.55]{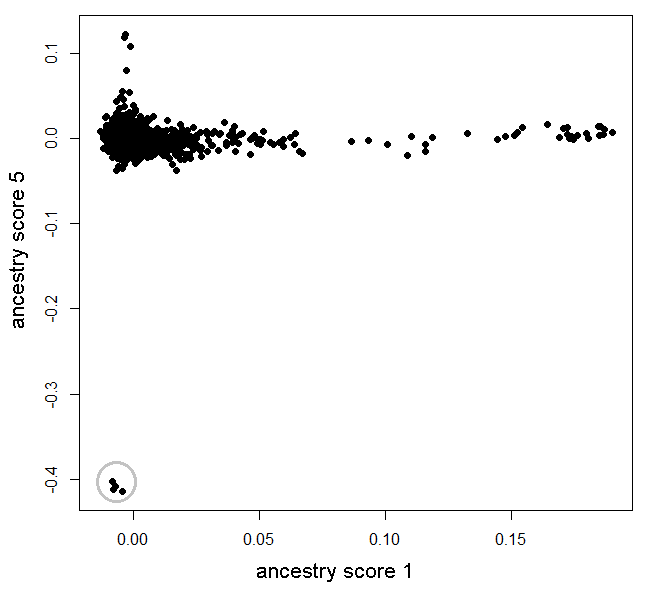}
\caption{Ancestry score (right singular vector) 5 vs. ancestry score 1 in a naive decomposition of the covariance matrix using all CF individuals. Membership in a family of size 4 (highlighted with a circle) is responsible
for most of the variation in ancestry score 5.}
\label{PC5vsPC1}
\end{center}
\end{figure}

Cystic Fibrosis (CF) is a recessive genetic lung disorder, caused by a mutation in the single gene {\it CFTR}. 
However, considerable genetic variation remains in the severity of disease, and evidence indicates this
variation is complex and influenced by numerous genes \cite{wright2011genome}.
Genotypes gathered by the North American CF Consortium are typical of a large-scale genomewide association
study (GWAS), with thousands of individuals and over 1 million genetic markers \cite{corvol2015genome}.
For covariate control, the eigenvectors are computed for a submatrix of the genotypes, after a ``thinning" process in which only an ancestry-informative
subset of markers which have low marker-marker correlation is retained \cite{patterson2006population}.
We illustrate the proposed methods using the dataset from the CF patients described as 'GWAS1' in  \cite{corvol2015genome},
with 21,205 thinned ancestry markers and 3444 individuals.
The data set includes 2546 singletons (unrelated to others) and 438 small families of siblings (417 sets of 2 individuals, 20 sets of 3, and 1 set of 4).  Figure \ref{PC5vsPC1} is a scatter plot of the fifth vs. the first ``ancestry scores" (right singular vectors for this example) from a naive analysis of all 3444 individuals (see Methods). 

Here the PC5 scores are driven largely  by membership in the family of size 4, rather than the ancestry substructure of interest.  Several additional top-ranked eigenvectors are also driven by family membership.  Accordingly, matrix projection methods have been proposed \cite{zhu2008unified}, in which singular value decomposition is performed on singletons, followed by projections for the remaining families.  However, this approach has been shown to produce shrunken projected scores for the family members \cite{lee2010convergence}. In \cite{conomos2015robust} , the PCAiR method was proposed to expand the set of individuals included in the SVD  to include a single individual from each family, resulting in improved performance. However, the question remains as to whether score for the remaining projected individuals will exhibit shrinkage, or if the methods can be further improved.

In contrast to previous efforts, in this paper we directly address the family covariance structures that complicate ancestry score calculation.  We introduce several novel approaches to account for the family-specific correlation structures in a single analysis, avoiding difficulties posed by standard
projection methods. Comparison via simulation and analysis of the CF data indicate that several of our approaches offer substantial improvements over existing methods, and are straightforward to implement. The analytic comparisons use both high dimensional geometry and the new device of smoothed individual scree plots.
The paper is organized as follows. In Section 2, we introduce the existing and proposed approaches. Section 3 describes performance criteria. Section 4 compares the methods using simulations.  Section 5 contains results from application to the real dataset. The Appendix contains details of the algorithms.


\section{Methods}
This paper discusses a large number of competing methods, and considerable notation is unavoidable.
To reduce confusion, we adopt uniform notation where possible.
We use $i=1,...,p$ to denote genetic markers (single nucleotide polymorphisms, SNPs), $j=1...,n$ to denote individuals (including families),
and typically $p>>n$. The individuals can be partitioned into singletons (${\mathcal S}$, unrelated to anyone else
in the dataset), and family members (${\mathcal F}$, related to at least one other individual), with respective sample sizes $n_{\mathcal S}$ and $n_{\mathcal F}$, so $n=n_{\mathcal S}+n_{\mathcal F}$.
The set ${\mathcal F}$ is partitioned into distinct families $\{{\mathcal F}_f\}$ of size $n_f$, $f=1,...,F$.
Let $G$ be the original $p \times n$ genotype matrix, with elements taking on the values 0, 1, or 2, typically coded as the number of minor alleles, and $\bar{g}_{i.}=\sum_{j=1}^n g_{ij}/n$, the mean for SNP $i$.
 The scaled $p \times n$ genotype matrix $X$ consists of elements 
$x_{ij}=(g_{ij}-\bar{g}_{i.})/\sqrt{\sum_{j^\prime}(g_{ij^\prime}-\bar{g}_{i.})^2/(n-1)}$,
so that $\sum_j x_{ij}=0$, $\sum_j x^2_{ij}=n-1$, for all $i=1,...,p$.  

\subsection{SVD and Eigen Decomposition}\label{SVD}
The ``naive" approach to handling the full dataset is to simply compute 
the singular value decomposition $X=UDV^T$, using the columns of $V$ as informative scores for
ancestry, in decreasing order of the singular values contained in the diagonal of $D$.
However, as Figure 1 showed, this approach can be highly influenced by family structure.
Other methods work with the matrix of sample covariances of the individuals, which for the full matrix $X$ is the $n \times n$ matrix
$M=\overline{X}^T \overline{X}/(p-1)$, where $\overline{X}$ is the column-centered version of $X$.
Eigen decomposition of $M$ provides eigenvectors that are nearly identical to the columns of
$V$.  Equivalently, a principal component (PC) decomposition provides PC scores that are
identical or nearly identical (depending on column-centering) to $V$. For 
ease of discussion we refer to the column output from the various methods simply as ``ancestry scores,"
except when further specificity is required.



\subsection{The Singleton Projection (SP) Method}\label{spm}
Singleton projection \cite{zhu2008unified} first computes the SVD $X_{\mathcal S}=U_{\mathcal S}D_{\mathcal S}V_{\mathcal S}^T$. Ancestry scores for the complete data are given as the columns of the $n\times n_{\mathcal S}$ matrix $\widetilde{V}_{SP}=X^T U_{\mathcal S} D_{\mathcal S}^{-1}$, as in practice no more than $n_{\mathcal S}$ ancestry
scores (PCs) will be used as covariates. Here and subsequently a tilde (``$\sim$") will signify a matrix or vector
that has been made robust to the effects of family relationships, and $\widetilde{V}$ with
a corresponding subscript will be used to denote the matrix of ancestry scores
for each method.
The singleton projection approach is easily implemented in popular software such as EIGENSTRAT \cite{price2006principal}.
By ignoring families in the initial step, singleton projection loses accuracy, with the family ancestry scores suffering 
from the shrinkage phenomenon described in \cite{lee2010convergence}, who also prescribed a bias-correction procedure to
correct the shrinkage. However, the bias-correction is a multi-step procedure whose performance has not been established for
a range of eigenvalues, and is not convenient for
collections of families of various sizes.

\subsection{PCAiR} \label{pcair}

To incorporate more information from the family data, PCAiR \cite{conomos2015robust} works with a set of unrelated individuals $\mathcal U$, where $\mathcal U$ includes the singletons plus a single
member from each family.  Thus $\mathcal U$ does not contain any related pairs, and we will use
$\mathcal R$ to denote the complementary set of related individuals not in $\mathcal U$.
The set $\mathcal U$ is not unique, and PCAiR attempts to identify and use a maximally-informative set.
The full approach \cite{conomos2015robust} involves genotype normalization differing slightly from our scaling, identification of family members using KING \cite{manichaikul2010robust}, and numerous matrix operations.
However, a careful reading shows that the essence of the approach is similar to singleton projection, using 
columns of $\widetilde{V}_{PCAiR}=X^T U_{\mathcal U}D_{\mathcal U}^{-1}$ as scores, where $U_{\mathcal U}, D_{\mathcal U}$ are obtained from the SVD
$X_{\mathcal U}=U_{\mathcal U}D_{\mathcal U}V_{\mathcal U}^T$. 
Although numerous ancestry estimation procedures have been proposed \cite{sankararaman2008estimating}, for the calculation of ancestry scores
using eigenvectors or principal components, the results in
\cite{conomos2015robust}
indicate that the PCAiR approach represents the current state of the art.  In Section \ref{Results}, for simple Gaussian simulations 
we use the algorithm coded above in $R$. However, for all simulated genotype datasets and the CF data, we
use the KING software and PCAiR code from \cite{conomos2015robust} as recommended. 

\subsection{Geometric Rotation / Family Whitening (FW)}\label{GRMW}

One critique of the existing approaches is that they do not use all of the data in computing the
 $U$ matrix, which corresponds to SNP loadings in a PC analysis.  A more direct approach
would be to include all of the data, but to first modify genotypes within families to reduce the family-specific
impact on SVD analysis.  Such modification is entirely for the purpose of stratification analysis -- the modified
genotypes are not intended to be used for trait association. We first describe the problem in geometric terms, to gain an understanding of the nature of the modification, and follow with the simple matrix operation analogue.  Our solution is to {\it rotate} the data to make individuals within a family orthogonal, performed within a plane such that the impact of the data rotation is otherwise minimal.
The approach is easiest to explain for a family of size 2, and the data for each individual is the scaled genotype
$p$-vector. Data vectors for first-degree relatives are expected to have a $60^\circ$ angle, corresponding to a genotype correlation of 0.5 (Appendix A).
We first find the mean vector of the two members, and then rotate each member away from the mean vector to a target angle of $45^\circ$.  This operation makes the new vectors orthogonal, which is approximately true for unrelated individuals.

In general, a family $f$ consists of $n_f$ individuals indexed by the set ${\mathcal F}_f$.
The target rotation angle $\theta_f$ is the same as the angle in $\mathbb{R}^{n_f}$ between each coordinate unit vector and the direction vector $(\frac{1}{\sqrt{n_f}},...,\frac{1}{\sqrt{n_f}})^T$, which is $\theta_{f}=arccos(\frac{1}{\sqrt{n_f}})$. 
For example, when $n_f=2$, $\theta_f=arccos \frac{1}{\sqrt{2}}=\pi/4$. 
Let $x_{.j}$ denote the data vector for individual $j$, with unit-length vector
$z_j=\frac{x_{.j}}{||x_{.j}||}$, where $||x_{.j}||$ is the length $\sqrt{\sum_i x_{ij}^2}$.
The mean vector $\bar{x}_{{\mathcal F}_f}$ is obtained by computing for each SNP $i$
$\bar{x}_{i{\mathcal F}_f}=\sum_{j\in {\mathcal F}_f} x_{ij}/n_f$, with unit vector $\bar{z}_{{\mathcal F}_f}=\frac{\bar{x}_{{\mathcal F}_f}}{||\bar{x}_{{\mathcal F}_f}||}$.
The unit length component of $z_j$ which is orthogonal to $\bar{z}_{\mathcal F}$ is 
$$\widetilde{z_{j}}=\frac{z_j-(z_j^T \bar{z}_{{\mathcal F}_f}) \bar{z}_{{\mathcal F}_f}}{||z_j-(z_j^T \bar{z}_{{\mathcal F}_f}) \bar{z}_{{\mathcal F}_f}||}.$$  
In the plane determined by $\bar{z}_{\mathcal F}$ and $\widetilde{z_{j}}$, the unit vector with angle $\theta_{f}$ to $\widetilde{z_{j}}$ is $\widetilde{\mu_j}=\cos(\theta_f) \widetilde{z_{j}} + \sin(\theta_{n_f}) \widetilde{z_{j}}$.  
The vector $\widetilde{x_{.j}}=\widetilde{\mu_j} ||x_{.j}||$ is the natural rescaling of $\widetilde{\mu_j}$, and
used as a replacement data vector for $x_{.j}$.
Finally the data vector for each family member is centered and rescaled to match the mean and variance of the original data. This rotation operation is conducted in succession for each family $f=1,...,F$, and SVD is applied to the new whitened data matrix.

Geometric rotation has a matrix operation interpretation, de-correlating the members of a family $f$ by an operation similar to classical multivariate sphering. Let $Z_{{\mathcal F}_f}$ be the $p\times n_f$ submatrix of scaled family genotype data, and $R_{{\mathcal F}_f}$ the corresponding (positive definite) $n_f\times n_f$ matrix of sample correlations. Then $\widetilde{Z}_{{\mathcal F}_f}=R_{{\mathcal F}_f}^{-1/2} Z_{{\mathcal F}_f}$ is a whitened matrix with identity correlation, and a final $\widetilde{X}_{{\mathcal F}_f}$ is obtained by recentering and scaling the columns of $\widetilde{Z}_{{\mathcal F}_f}$ to match the mean and variance of the original $X_{{\mathcal F}_f}$. Finally, the columns of singletons and newly whitened family data are combined
into $\underset{p\times n}{\widetilde{X}}=[\underset{p\times n_{\mathcal S}}{X_{\mathcal S}},\underset{p\times n_{\mathcal F}}{\widetilde{X}_{\mathcal F}}]$, and the ancestry scores are $\widetilde{V}_{FW}$ from the SVD 
$\widetilde{X}=\widetilde{U}\widetilde{D}\widetilde{V}_{FW}^T$.

\begin{figure}
\begin{center}
\includegraphics[scale=0.55]{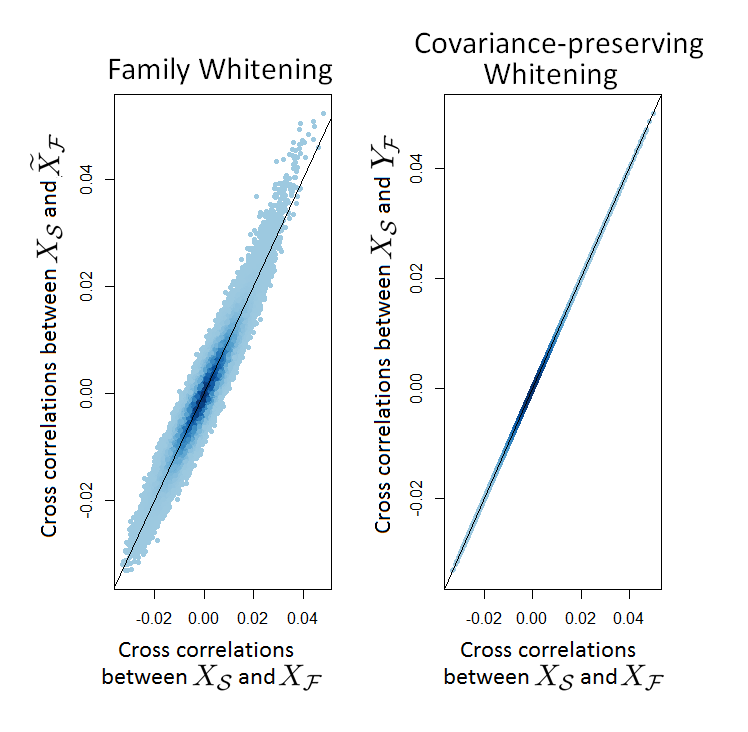}
\caption{Cross-correlations between genotype vectors of set $\mathcal S$ vs. $\mathcal F$. On each axis, a point represents a correlation
between an individual in $\mathcal S$ to an individual in $\mathcal F$, for a total of 2546$\times$ 898 points.
Left panel: Cross correlations of 
$X_{\mathcal S} \times \widetilde{X}_{\mathcal F}$ vs. cross correlations of $X_{\mathcal S} \times {X}_{\mathcal F}$ show
modest deviation. 
Right panel: Cross correlations of 
$X_{\mathcal S} \times Y_{\mathcal F}$ vs. cross correlations of $X_{\mathcal S} \times {X}_{\mathcal F}$ show that the goal of covariance-preserving whitening is achieved.}
\label{correlation_change}
\end{center}
\end{figure}

In practice, geometric rotation and matrix whitening of the family are nearly identical, with slight differences due to handling of column centering, and the matrix approach is used subsequently.
Figure \ref{correlation_change} (left panel) shows the result of family whitening in the CF dataset, in terms of the correlation of columns of $\widetilde{X}_{\mathcal F}$ compared
to those of  $X_{\mathcal S}$. This shows that the family whitening operation introduces some perturbation of the correlation structure. We will return to this issue below.


\subsection{Matrix Substitution (MS)}\label{MS}
The within-family rotation/whitening method reduces the strong impact of families in stratification analysis.
However, as seen in the left panel of Fig \ref{correlation_change}, the approach is not ideal, as we observed that the whitening operation also affects the covariance of family members with the remaining sample. A question arises as to whether within-family data can be orthogonalized without changing the covariance relationship of these family members to the remaining individuals. Before answering this question, we consider the following ``direct" approach. As noted, ancestry scores can be obtained directly from a covariance matrix \cite{frudakis2003classifier}, and we propose modifying the sample covariance matrix $M=\overline{X}^T \overline{X}/(p-1)$.  We construct a matrix $\widetilde M$ with entries ${\widetilde m}_{j_1 j_2}=${\it median entry in $M$} if $j_1 \ne j_2$ and $j_1$ and $j_2$ belong to the same family, and ${\widetilde m}_{j_1 j_2}=m_{j_1 j_2}$ otherwise.  
Co-family members are typically a small fraction of the pairs of individuals, and so $M$ and $\widetilde M$ differ in only a small
fraction of elements.

Family membership could be inferred by KING \cite{manichaikul2010robust} or other purpose-built software.  However, a simple screening method for first-degree relationships is also effective, identifying pairs of individuals $j_1, j_2$ such that ${\rm corr}(x_{.j_1},x_{.j_2})>\eta$, and $\eta=0.4$ identifies paired family members with high sensitivity and specificity (see Appendix A).  Following matrix substitution, we compute 
$\widetilde{V}_{MS}$ as the eigenvectors of $\widetilde{M}$.

\subsection{Covariance-Preserving Whitening (CPW)}\label{CPW}

Although the matrix substitution approach is appealing, it does not provide whitened genotype data, which might be useful for other purposes, such as  analyses of subsets of individuals or for careful investigation of marker-marker correlation \cite{lake2000family}.  Here we describe an approach to modify the genotypes within families so that the final covariance matrix equals the modified covariance matrix $\widetilde M$ described above, and families are orthogonalized while retaining their covariance with the remaining sample.
The goal here is to find an $n\times n$ matrix $B$ such that $Y=X B^T$ and $\frac{1}{p-1}Y^TY=\frac{1}{p-1} B\bar{X}^T\bar{X}B^T=\widetilde{M}$, where the entire sample, including all families, is handled at once. There are multiple possible solutions, but it is appealing to add the constraint that only  family members be modified, as singletons do not contribute to the problem of ``spurious" ancestry scores.   We assume that the columns of $X$ are arranged with singletons $\mathcal S$ followed by families $\mathcal F$. We then divide $M$ (defined above) and $B^T$ into submatrices as follows,
 $$
\underset{n \times n} M=
\begin{bmatrix}
  \underset{(n-n_{\mathcal F})\times (n-n_{\mathcal F})} {M_{11}}     & \underset{(n-n_{\mathcal F})\times n_{\mathcal F}}{M_{12}} \\
  \underset{n_{\mathcal F} \times (n-n_{\mathcal F})}{M_{21}}   &   \underset{n_{\mathcal F}\times n_{\mathcal F}}{M_{22}}\\
\end{bmatrix},~~
\underset{n \times n} B^T=
\begin{bmatrix}
   \underset{(n-n_{\mathcal F})\times (n-n_{\mathcal F})}{I_{n-n_{\mathcal F}}}       & \underset{(n-n_{\mathcal F})\times n_{\mathcal F}}{C} \\
   \underset{n_{\mathcal F} \times (n-n_{\mathcal F})}{0}  &  \underset{n_{\mathcal F}\times n_{\mathcal F}}{D}
\end{bmatrix},
$$
where $n_{\mathcal F}$ individuals belong to the all-families set $\mathcal F$, and 
$I_{n-n_{\mathcal F}}$ denotes an $(n-n_{\mathcal F})\times (n-n_{\mathcal F})$ identity matrix.
Note that $\widetilde{M}$ differs from $M$ in only the co-family pairs of the lower right submatrix,
and we will use $\widetilde{M}_{22}$ to denote the corresponding $n_{\mathcal F}\times n_{\mathcal F}$ lower right submatrix of $\widetilde{M}$.
The form of $B^T$, with the identity submatrix operating on the singletons in $Y=X B^T$, achieves the desired 
constraint that singletons be unchanged. $C$ and $D$ are unknown matrices, to be solved for.
We show in Appendix B that the solution for full-rank $X$ is 
$$
C=M_{11}^{-1} M_{12} (I_{n-n_{\mathcal F}}-D), ~~D= (M_{22}-S)^{-1/2}(\widetilde{M}_{22}-S)^{1/2},$$
where $S=M_{12}^T (M_{11}^{-1})^T M_{12}=M_{21} M_{11}^{-1} M_{12}$,
with a slight modification to account for our situation that $X$ has rank $n-1$.
For $Y=X B^T$, ancestry scores are obtained as $\widetilde{V}_{CPW}$ in the SVD $Y=\widetilde{U}\widetilde{D}\widetilde{V}_{CPW}^T$.
Figure \ref{correlation_change}  (right panel) shows the result of 
applying covariance-preserving whitening to the CF data. The plot shows that, for the new matrix $Y$, cross-correlations of families $\mathcal F$ to singletons $\mathcal S$ have indeed been preserved from the original $X$.  In fact even the correlations between members of different families have been preserved (not shown).

\subsection{Family Average (FA) Projection}\label{FAP}
A  concern with the PCAiR projection method of Section \ref{pcair} is that
only a single member is used from each family. We consider the potential improvement of using
the mean vector for each family, instead of a single representative member,
to obtain loadings.  Specifically, for family $f$ indexed by ${\mathcal F}_f$, we compute a new data vector $\widehat{x}_{.f}=\bar{z}_{{\mathcal F}_f}(\sum_{j\in{\mathcal F}_f} || x_{.j}||/n_f)$,
where $\bar{z}_{{\mathcal F}_f}$ is the unit-length family mean vector from \ref{GRMW}. Multiplication by the family average length ensures that $\widehat{x}_{.f}$ has a ``typical" length --  otherwise the variance contribution from the family mean vector would be much smaller than for an individual.  We construct a new matrix of singletons combined columnwise with the $F$ rescaled family averages, $\underset{p\times (n_{\mathcal S}+F)}{X_{\mathcal A}}=[\underset{p\times n_{\mathcal S}}{X_{\mathcal S}}, \underset{p\times F}{\widehat{X}}]$, and compute the SVD
$X_{\mathcal A}=U_{\mathcal A}D_{\mathcal A}V_{\mathcal A}^T$. Finally, the projected ancestry
scores are computed for all individuals, as the columns of $\widetilde{V}_{FA}=X^T U_{\mathcal A}D_{\mathcal A}^{-1}$.

\section{Criteria for evaluation}
Here we describe several criteria to evaluate the performance of ancestry score calculations.
The first two criteria reflect the ability to discriminate among known (by simulation) population strata, while
providing family ancestry scores that are comparable to those from singletons.
The third criterion, which can be assessed with real data, measures the tendency for ancestry scores to remain stable 
for an individual who belongs to a family, depending on whether the individual's family members are also included in the analysis.
Finally, we end this section by introducing the ``individual scree plot," a novel visualization tool to provide insight
into the behavior of ancestry scores.


\subsection{The Standardized Within class Sum of Squares (SWISS)  Criterion }\label{SWISS}
The ancestry scores are columns of a matrix $\widetilde V$,
where each entry $v_{jl}$ is the $l$th ancestry score
for individual $j$.
We assume the population is partitioned into $K$ distinct strata (ancestry subgroups), and the indices for individuals belonging to the $k$th subgroup are $j\in {\Omega}_k$, $k=1,...,K$.
The SWISS criterion  \cite{cabanski2010swiss}
is similar to $1-R^2$ in analysis of variance, with strata as factor levels. 
For the $l$th ancestry score, let $\overline{ \overline{v}}_{.l}$ be the overall mean and $\overline{ v}_{\Omega_k l}$ be the  mean for the $k$th stratum. 
The SWISS  value for the $l$th ancestry score is
$$SWISS_l=\frac{\sum_{k=1}^K \sum_{j\in {\Omega}_{k}}(v_{jl}-\overline{v}_{\Omega_k l})^2}{\sum_{j} (v_{jl}-\overline{ \overline{v}}_{.l})^2}.$$ 
We average across the first 5 $SWISS_l$ values to compute an overall SWISS score.
Smaller SWISS values indicate a higher ability to discriminate among strata. 


\subsection{The Relateds Square Error  (RSE) Criterion}\label{error1}
 
Most of the methods described in this paper use a partition into
family members $\mathcal F$ vs. singletons $\mathcal S$.
An important performance aspect that is not well captured by SWISS is the tendency for the family members
to exhibit reduced variation in the ancestry scores.
We introduce a finer-grained measure of the tendency for ancestry scores of family members to overlap their singleton counterparts, calculated within each stratum before summarizing.

For each stratum $k$, we further partition ${\Omega}_k$ into ${\Omega}_{k,{\mathcal F}}$ and ${\Omega}_{k,{\mathcal {\mathcal S}}}$, corresponding to family members and singletons within the stratum, of sizes $n_{k,{\mathcal F}}$ and $n_{k,{\mathcal S}}$. 
Let  $\overline{ v}_{\Omega_{k,{\mathcal S} l}}$ denote the average of the $l$th ancestry scores for individuals in ${\Omega}_{k,{\mathcal S}}$. For the $l$th ancestry score, we compute the Relateds Squared Error (RSE),
$$
RSE_l=\sqrt{\frac{\sum_{k=1}^K \sum_{i \in {\Omega}_{k,{\mathcal F}}} (v_{jl}-\overline{ v}_{\Omega_{k,{\mathcal S}} l})^2/(n_{k,{\mathcal F}}-1)}{\sum_{k=1}^K \sum_{i \in {\Omega}_{k,{\mathcal S}}} (v_{jl}-\overline{ v}_{\Omega_{k,{\mathcal S}} l})^2/(n_{k,{\mathcal S}}-1)  }} .$$
In other words, for both family members and singletons, we compute the average squared deviation from the mean
of singletons. For a method that performs well, projected family members will behave similarly to
singletons,
and $RSE_l$ will be near 1.0.
We average the first 5 $RSE_l$ values to obtain an overall RSE.
For PCAiR, we compute the RSE using $\mathcal U$ and $\mathcal R$ instead of $\mathcal S$ and $\mathcal F$, respectively.

\subsection{An instability index}

The criteria above require knowledge of the true population strata. 
Here we describe a performance criterion based on {\it stability} of the eigenvector values for family members, as compared to an internally-computed standard.  It can be performed for real data, and thus applies to admixed settings where individuals cannot be cleanly classified into discrete strata. We will let $\underset{n\times n}{W}$ denote a ``gold standard" ancestry matrix to be used subsequently,
and $\underset{n\times n}{Q}$ a comparison matrix, and for both matrices the columns are arranged in the same order as $X$.

Suppose we wish to compute ancestry scores for an individual $j$ who belongs to a family.
One approach, robust to family structure, is to combine $j$ with the singletons, computing
$\underset{p\times (n_{{\mathcal S}+1})}{X_{{\mathcal S}\cup j}}=UD\underset{(n_{{\mathcal S}+1})\times (n_{{\mathcal S}+1})}{V^T}$.   As $j$ is unrelated to $\mathcal S$, we will use the last column of $V$ as the $j$th column of $W$, i.e. $w_{.j}=v_{. (n_{\mathcal S}+1)}$.
We perform this procedure in succession for all $j\in {\mathcal F}$ to populate the family ($\mathcal F$) columns of $W$. 
Alternately, 
we populate the $\mathcal F$ columns of $Q$ by performing,
 for each $f\in \mathcal{F}$, the family-robust methods described in
this paper, applied for each $f$ using the genotype data for $S\cup \mathcal{F}_f$.  In other words, $W$ is computed by combining
each family member with $\mathcal S$ one at a time, while $Q$ is computed by combining each {\it family} with $\mathcal S$. 
We consider $W$ as the gold standard, because it is computed using only unrelated individuals in each step.  For an ancestry
method that is robust to family structure, we expect $Q$ to be similar to $W$.
The instability index for the $l$th ancestry score is $instability_l=\sum_{j\in {\mathcal F}} (q_{jl}-w_{jl})^2/\sum_{j\in {\mathcal F}}q_{jl}^2$, with an ideal value of zero. 

\subsection{Individual Scree Plots} \label{three}
Scree plots \cite{cattell1966scree} are a useful method to visualize the relative importance of eigenvectors and PCs.
Here we take the scree plot in a new direction, by studying the corresponding plot for each individual, i.e. studying the squares of the projections of each individual.  For the SVD $X=UDV^T$, these projections are $(X^TU)^2=(VD)^2$, and the column sums of $(VD)^2$ are the squared singular values of $X$. These values essentially correspond to
principal component variance values, which are also used in overall scree plots.
Accordingly, for the robust ancestry methods described in this paper, we use rows of $(\widetilde{V}\widetilde{D})^2$ as {\it individual} scree values, reflecting the contribution of each
individual to the overall influence of each ancestry score. 
The individual scree values are noisy and cover several orders of magnitude, so we
plot them on the ${\rm log}_{10}$ scale and perform loess smoothing to discern important patterns.

\section{Genotype Simulation Methods and Settings}

Much of the behavior of the various methods can be understood largely in terms of covariance patterns, and are not unique to discrete
genotype data.  This is seen using  idealized Gaussian simulations in the supplementary material.  Another informative set of simulations more directly reflects the special origins of genotype data, which is studied next.


\subsection{Simulation of genotypes and family sibships}

Appendix C describes our procedure for realistic simulation of founder genotype data for $K$ population strata, following the Balding-Nichols model. The model uses modest serial correlation of successive markers of approximately 0.2 in blocks of 20 markers, 20,000 markers in total, and matches the
allele frequencies in the CF data.
To simulate a family sibship of size $n_f$, we followed a realistic autosomal recombination model.
First, we generated enough singletons within each subpopulation so that parents could be simulated and then discarded. For each family, from the singletons we randomly selected two parents at random from a stratum (subpopulation) without replacement.
Artificial grandparental haplotype genomes were generated for each parent by randomly dividing the alleles. Children were then simulated using an artificial recombination process, with recombinations in each parent simulated as a geometric random variable for successive SNPs, at a rate such that on average 30 recombinations occurred per meiosis. For each family, the $n_f$ children were simulated independently from the same parental pair.

\subsection{Balanced vs. unbalanced families per subpopulation}\label{balance}
For the balanced simulations, we generated $K=5$ subpopulations using the approach above.
Sibships of $n_f=3$ were simulated such that the proportion of individuals belonging to families $prop$ was the same in each subpopulation. The total sample sizes used were $n=\{500, 1000, 2000\}$, with
family proportions $prop=\{0.2, 0.5, 0.8\}$, respectively, and the total number of families was $n (prop) /3$.

For the unbalanced simulations, again 5 subpopulations were simulated with total $n=\{500, 1000, 2000\}$. However, all of the families, again with $n_f=3$, were simulated from a single subpopulation, such that 20\% of the total sample size belonged to these sibships. This scenario was intentionally extreme, to determine the robustness of various methods for handling families.

\section{Results}\label{Results}

The supplementary file shows results for Gaussian simulations for $p=10,000$ and varying proportions
with unrelated individuals and ``family pairs" that have correlation 0.5. 
 The number of strata was $K=3$, so
two ancestry scores are sufficient to capture the relationships, and visual impressions can be formed.
Supplementary Figure 1 illustrates that singleton projection results in extreme
shrinkage of projected family members $\mathcal F$, while the PCAiR algorithm results in modest shrinkage of the individuals in $\mathcal R$.
Matrix whitening shows modest shrinkage for $\mathcal F$, while the remaining novel methods all show good and similar performance.  
For highly unbalanced data with all families coming from a single subpopulation (Supplementary Figure 2), the conclusions are similar, although shrinkage
is less extreme due to a higher overall proportion of singletons. The findings are sensible, reflecting the simple fact that inclusion of
an individual when computing loadings results in better performance. For family whitening, the change in cross correlations with
individuals outside the family results in family shrinkage.

\subsection{SWISS and RSE criteria for simulated genotypes}

\begin{figure}
\begin{center}
\includegraphics[scale=0.45]{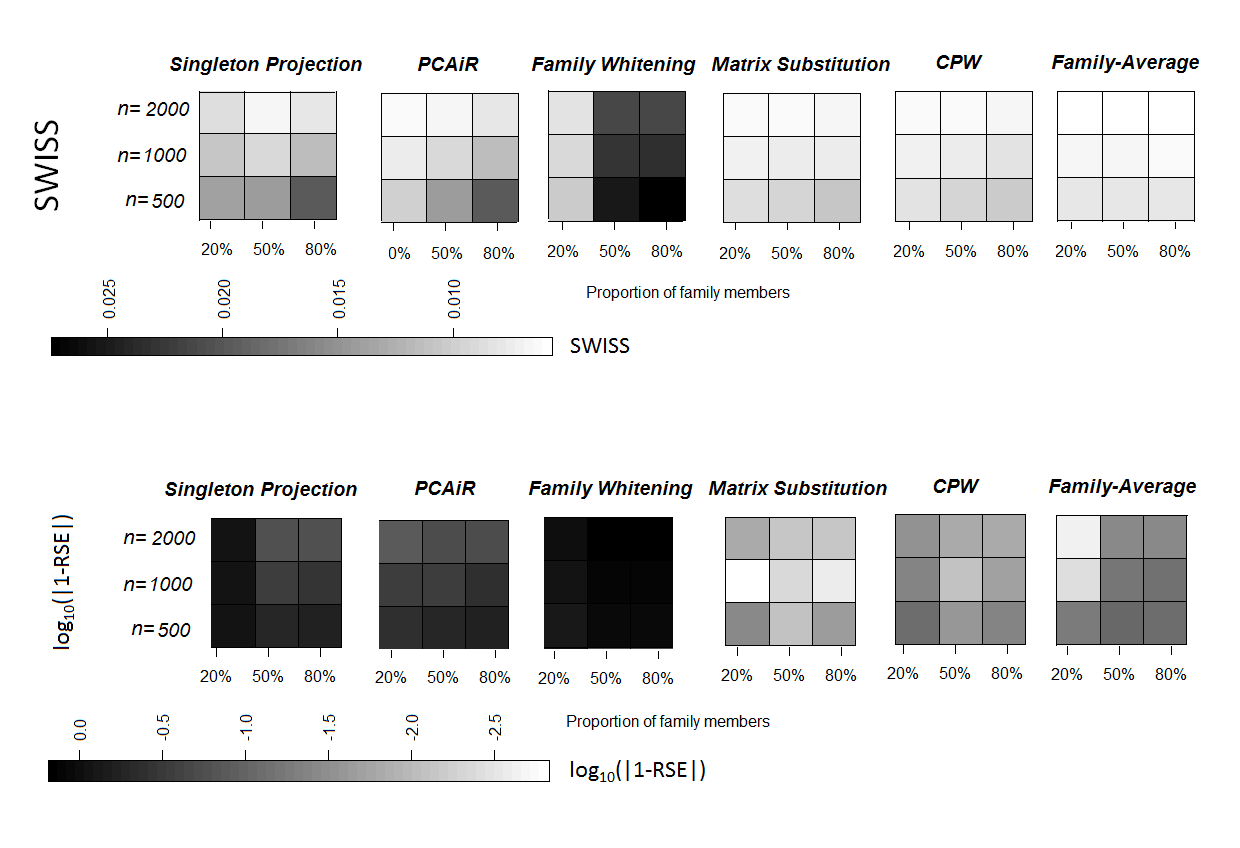}
\caption{Heatmap for SWISS and RSE performance, for the balanced simulations with the same proportion of family members in each of $K=5$ subpopulation strata.
 }
\label{fig:balanced}
\end{center}
\end{figure}

Figure \ref{fig:balanced} depicts results in heatmap form for our genotype simulations in which the proportion of families is balanced across
the 5 strata.   For the SWISS criterion, Matrix Substitution, CPW, and Family Averaging appear to perform similarly and somewhat better than PCAiR. For the RSE criterion, differences are more noticeable, and again Matrix Substitution, CPW, and Family Averaging perform the best. 
For large samples ($n=2000$) and a modest proportion of family members (20\%), family averaging performs best. Family whitening performs  poorly.

\begin{figure}
\begin{center}
\includegraphics[scale=0.45]{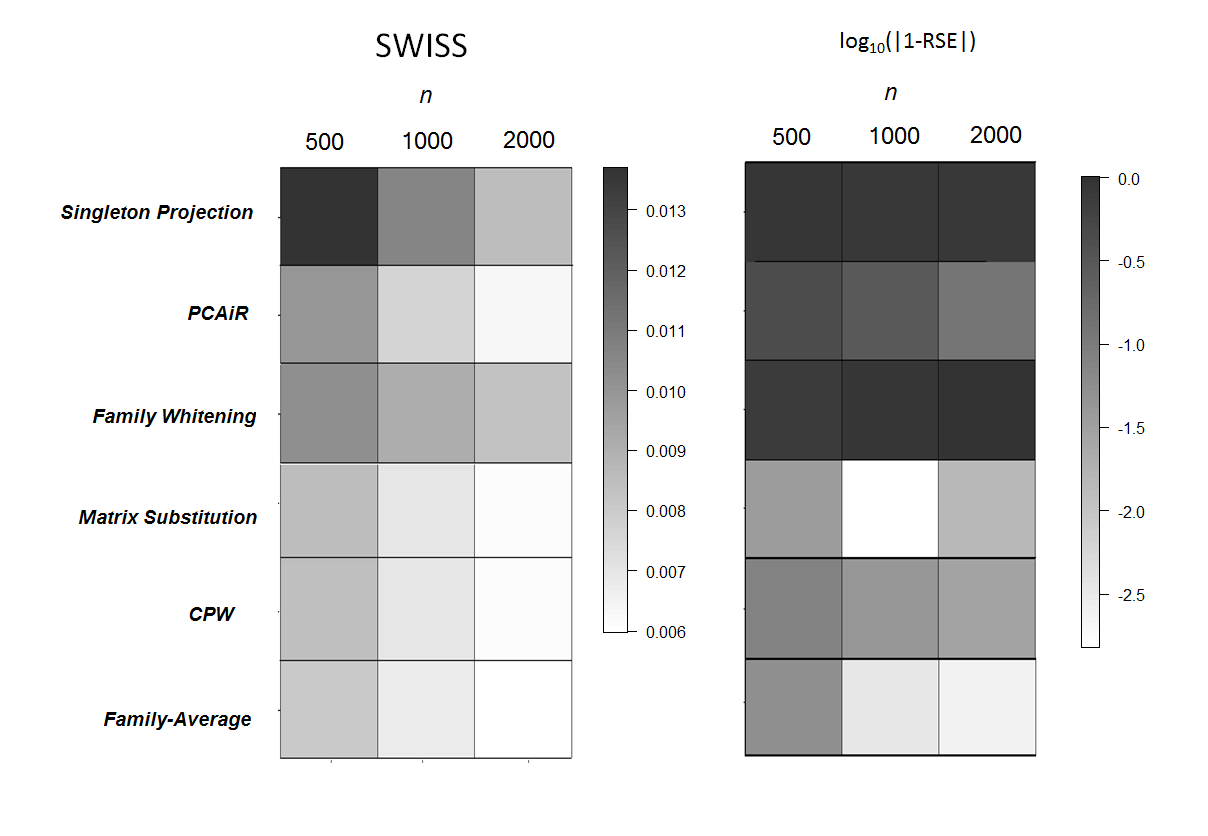}
\caption{Heatmap for SWISS and RSE performance, for the unbalanced simulations with 20\% of the sample consisting of family members in a single stratum.}
\label{fig:unbalanced}
\end{center}
\end{figure}

Figure \ref{fig:unbalanced} shows the performance of the methods under the unbalanced genotype simulation with
20\% of individuals belonging to families, from a single stratum. 
The left panel shows the SWISS performance, for which family averaging offers a slight improvement over matrix substitution and CPW, followed by PCAiR.  For the RSE criterion, ranking of methods is similar, with family averaging performing especially well for larger sample sizes. 
As expected, performance generally improves with increasing sample size.

We next applied the methods to the CF dataset, using the instability index approach described earlier. To do so, we 
first performed 898 separate analyses of ${\mathcal S}\cup j$ for each $j\in{\mathcal F}$.  We then performed 438 analyses
of ${\mathcal S}\cup {\mathcal F}_f$ for each $f=1,...,438$, and compared the two sets of analyses using the instability index,
for each of the first 6 ancestry scores.

The three scatterplots in Figure \ref{fig:instability} show the results for the first and second ancestry scores using covariance matrix
eigen-decomposition and a single family with two siblings.  The A and B panels show the position of ancestry scores when the two siblings 
are analyzed separately. Panel C shows the  the results for the entire family after matrix substitution, overplotted with the values from earlier panels, showing that they have changed little. The D panel shows the stability index values for ancestry scores 1-6 (which are all the scores clearly meeting significance thresholds, \cite{corvol2015genome}) 
and the various methods.   Singleton projection and family whitening performed much more poorly, and are not shown.
As expected, matrix substitution and covariance-preserving whitening were nearly identical, and performed similarly to PCAiR for the first 4 ancestry scores.  However, for ancestry scores 5-6, PCAiR showed much higher values of the instability index.  Family averaging showed considerably lower instability for eigenvectors 1-4.

\begin{figure}
\begin{center}
\includegraphics[scale=0.58]{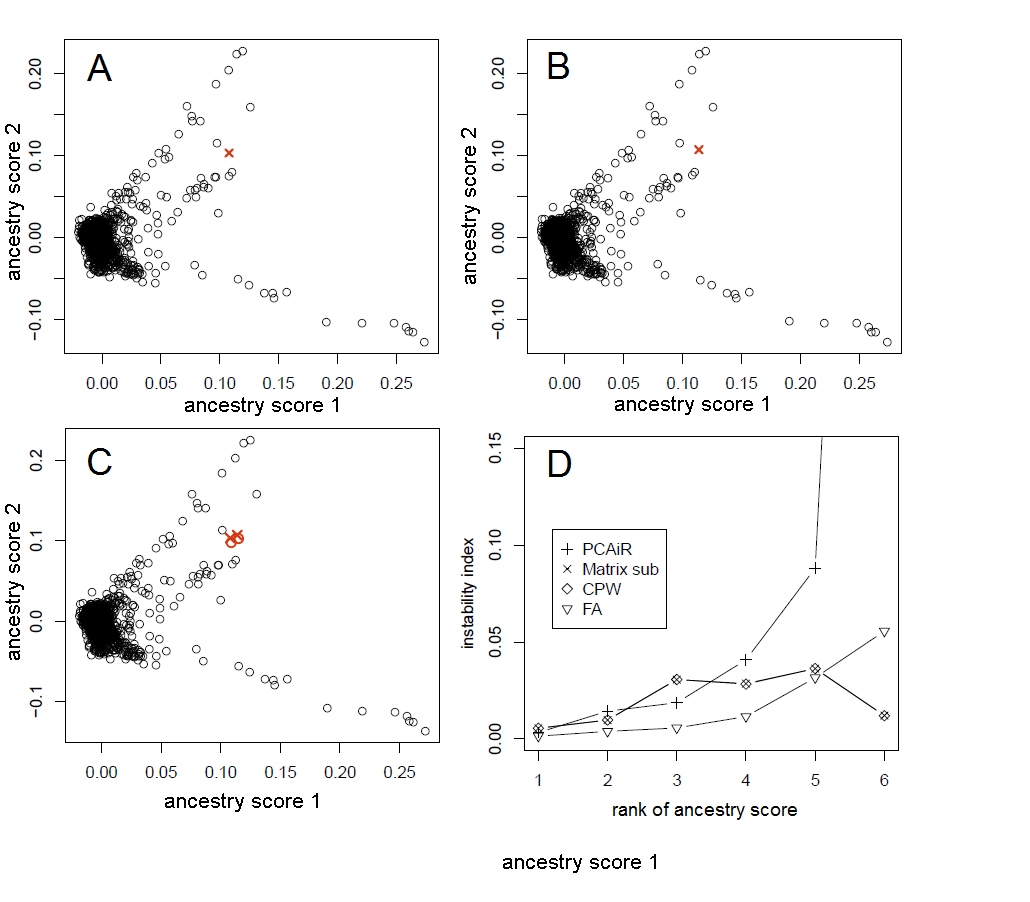}
\caption{Illustration of the instability index for the CF dataset. A) Ancestry scores (eigenvectors of the covariance matrix) for all singletons plus the first sib in a family, marked as a red ``X". 
B) Ancestry scores for all singletons plus the {\it second} sib in the family.  C) Ancestry scores for matrix substitution, with the two individuals shown as circles, overplotted with values computed in A and B.
D) The instability index for each method, providing a summary for each ancestry score across the 898 family members.
 }
\label{fig:instabilityfred}
\end{center}
\end{figure}

\subsection{Individual Scree Plot Results}

\begin{figure}
\begin{center}
\includegraphics[scale=0.48]{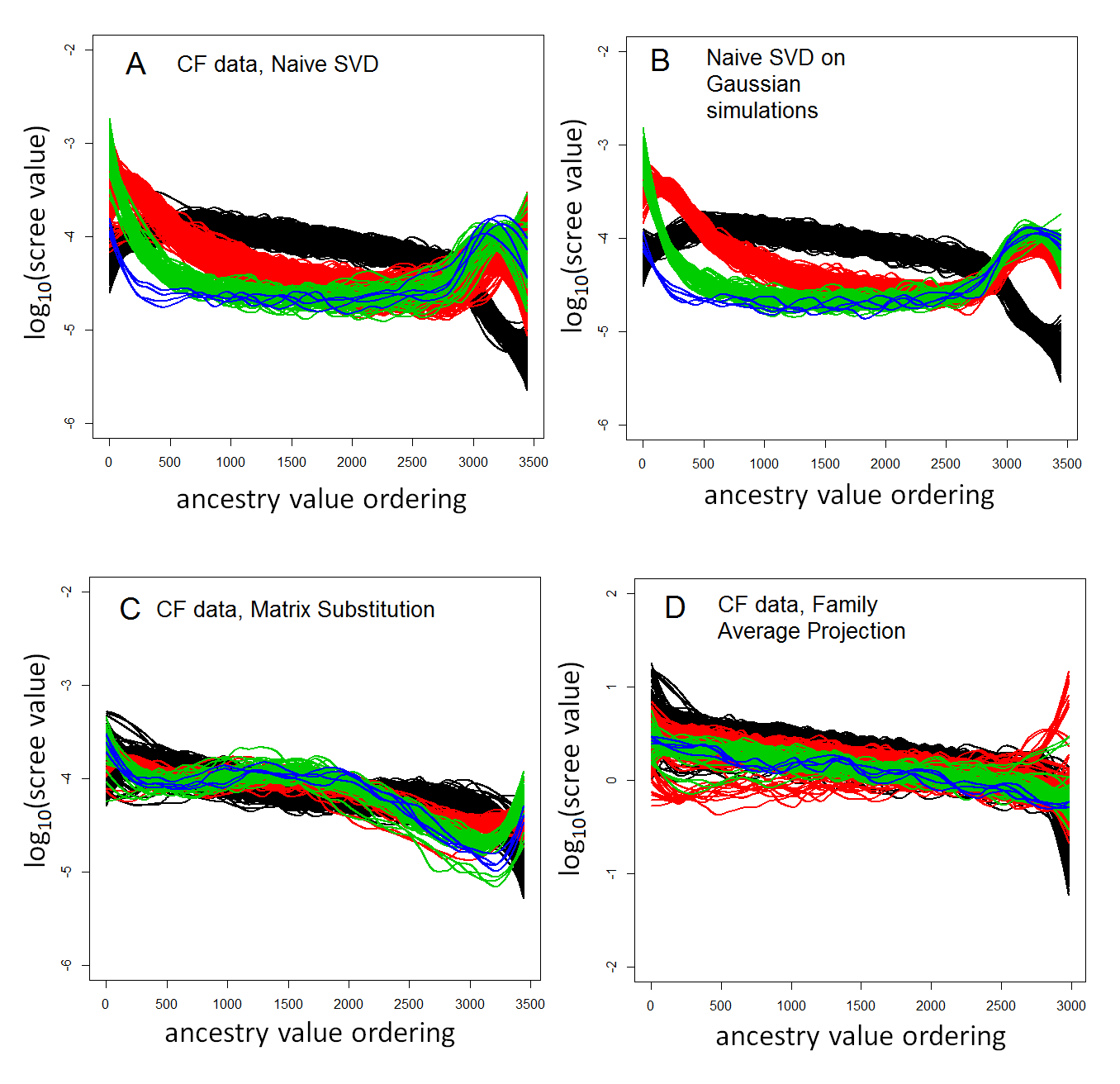}
\caption{Individual scree plots for several methods. Black curves are for the singletons; red curves show members of families of size 2; green curves are for families of size 3 and blue curves are for the family of size 4. A) Individual scree curves for the full CF data using naive ancestry analysis, with all individuals included; B) The plot for simulated Gaussian data with the same family structure as the CF data; C) The plot for the full CF data using matrix substitution, showing that the ``removal" of family effects persists through most of the ancestry values; D) The plot using the family average approach suggests further improved removal of family effects.
 }
\label{fig:scree}
\end{center}
\end{figure}

Overall, the simulations and real data showed that the novel methods (except family whitening) dominate PCAiR and singleton projection.
To gain further insight into the properties of the various methods, we examined the individual scree plots for
the full CF dataset (Figure \ref{fig:scree}), with curves colored according to the size of the family that each individual belongs to.
Panel A of Figure \ref{fig:scree} shows the individual scree curves for the naive analysis, which simply applies SVD to the full dataset without
regard to the presence of families. The colored curves (red, green, blue) show these curves for family members from families of various sizes (2, 3, 4, respectively).  Although the individual scores are highly variable (see Supplementary Figure 3), after smoothing the patterns are broadly consistent. Family members have higher values for the first components, because they tend to drive the highest-ranked ancestry scores in a naive analysis.  Family members tend to have lower curves for the middle scores, because these ancestry directions are driven by the non-family members (as expected).  Family members again have larger values for the last ancestry components, because these directions are driven by family component direction vectors that are {\it orthogonal} to the dominant family direction. 

To carefully check these interpretations, we performed a simulation study using Gaussian data, with the approach described in the Supplement, and  the numbers of each family type ($n_f=2,3,4$) matching the real CF data (panel B of Figure \ref{fig:scree}). The family patterns are very similar, although with somewhat less scatter, indicating that the geometric interpretations of these patterns are correct. 
Panel C shows the individual scree curves for matrix substitution, for which the curves of family members more closely overlap those of singletons.
However, the curves for families of size 3 and 4 remain distinctive, as matrix substitution does not fully eliminate the effect of high
correlation between family members.  Panel D shows that the family average method achieves more general overlap of scree curves among the individuals.

\section{Summary and Conclusions}

With the CF dataset as a motivating example, we have introduced several new methods to obtain family-robust informative ancestry scores in
genetic stratification analysis.  Several of the methods offer improvements over the current standard, and yet are quite simple to perform using standard matrix operations, which are available as {\it R} code from the authors.  Our careful genotype simulations and analysis of the CF data support the general motivating discussion in the supplement. In particular, both singleton projection and (to a lesser extent) PCAiR suffer from shrinkage due to the exclusion of individuals when computing loadings.

Among the new methods, family average projection appears to perform better than matrix substitution and covariance-preserving whitening, although the improvement is slight. The matrix substitution method has a potential advantage in that it relies only on the $n\times n$ covariance matrix, which is typically much smaller than the original genotype dataset. Covariance-preserving whitening may be appealing if the resulting whitened matrix is to be used in further investigations of linkage disequilibrium structure, or perhaps in substructure analysis of individual chromosomes. 

Alternative stratification control methods have included case-control modeling based on stratification scores \cite{epstein2007simple}, which
rely importantly on high-dimensional data summaries as part of the modeling procedure. Thus we foresee the methods described herein as providing useful ancestry scores for subsequent careful modeling of disease risk in combined sets of related and unrelated individuals.


\section{Appendix}
\subsection{Appendix A. Genotype correlation between first-degree relatives.} 
Standard results for shared genotype probabilities for related individuals are expressed in terms of kinship coefficients and identity-by-descent probabilities.  Here we clarify, as is needed for this paper, the {\it correlation} of genotypes between first-degree relatives.
We focus on siblings, although a slight modification of the argument applies to parent-child relationships.
Let $q$ denote the minor allele frequency, and a pair of siblings have random genotypes $g_1$ and $g_2$, with
means $2q$ and variances $2q(1-q)$. We have
\begin{eqnarray*}\label{correlation}
corr(g_1, g_2)&=&\frac{E(g_1 g_2)-E(g_1) E(g_2)}{SD(g_1)SD(g_2)}=\frac{E(g_1 g_2)-(2q)^2}{2q(1-q)}.
\end{eqnarray*}  
The identity-by-descent (IBD) outcomes determine $E(g_1 g_2)$. For IBD=0, $E(g_1 g_2|IBD=0)=(2q)^2.$ 
Also,  $E(g_1g_2|IBD=2)=E(g_1^2)=2({\rm var}(g_1)+E(g_1)^2)=2q(1-q)+(2q)^2.$ If IBD=1, without loss of generality, we assume the shared allele comes from the mother. We use $a_m$ to denote the allele from the mother and $a_f$ from the father. Then $E(g_1g_2)=E((a_{m1}+a_{f1})(a_{m1}+a_{f2}))=E(a_{m1}^2+a_{f1}a_{m1}+a_{m1}a_{f2}+a_{f1}a_{f2})=q+3q^2.$ Therefore
\begin{eqnarray*}
E(g_1 g_2) & = & E(E(g_1 g_2|IBD))\\
 & = & \frac{1}{4} E(g_1g_2|IBD=0) + \frac{1}{2} E(g_1 g_2|IBD=1)+\frac{1}{2}E(g_1 g_2|IBD=2)\\
& = & \frac{1}{4}(2q)^2 + \frac{1}{2} (2q(1-q)+(2q)^2)+\frac{1}{4}(q+3q^2)
\end{eqnarray*}
and plugging in to the correlation gives  0.5, regardless of $q$. 

\subsection{Appendix B. The Covariance-Preserving Whitening Solution} 
We have
$$
A M A^T=
\begin{bmatrix}
  I_{n-n_{\mathcal F}}       & 0^T  \\
   C^T       &  D^T \\
\end{bmatrix}
\begin{bmatrix}
  M_{11}      & M_{12} \\
 M_{21}    &  M_{22}\\
\end{bmatrix}
\begin{bmatrix}
   I_{n-n_{\mathcal F}}       & C  \\
  0      &  D \\
\end{bmatrix}
$$
$$=
\begin{bmatrix}
  M_{11}       & M_{11}C +M_{12}D \\
C^T M_{11} +D^T M_{21}      & \underbrace{ C^T M_{11} C}_{a}+ \underbrace{C^T M_{12} D}_{b} +\underbrace{ D^T M_{21} C}_{c}+ \underbrace{D^TM_{22} D}_{d}\\
\end{bmatrix}$$
$$
=\begin{bmatrix}
  M_{11}      & M_{12} \\
 M_{21}    &  \tilde{M}_{22}\\
\end{bmatrix}.
$$
Comparing the last two expressions provides two equations in the two unknowns $C$ and $D$.  From the upper right, we have
$M_{11} C+M_{12} D=M_{12}$, which implies $C=M_{11}^{-1} M_{12} (I_2-D)$ (the lower left is the same equation written in transpose form).  The lower right requires a bit more effort. We consider each of the four terms separately, plugging in the solution for $C$ from above, giving
$$ a=C^T M_{11} C=(M_{11}^{-1} M_{12} (I_2-D))^T M_{11} M_{11}^{-1} M_{12} (I_2-D)$$
$$=(I_2-D)^T M_{12}^T (M_{11}^{-1})^T M_{12} (I_2-D)=(I_2-D)^T S (I_2-D),$$
where $S=M_{12}^T (M_{11}^{-1})^T M_{12}=M_{21} M_{11}^{-1} M_{12}$.
$$ b=C^T M_{12} D=(I_2-D)^T M_{12}^T(M_{11}^{-1})^T M_{12} D= (I_2-D)^T S D$$
$$ c=D^T M_{21} C=D^T M_{21} M_{11}^{-1} M_{12} (I_2-D)=D^T S (I_2-D),$$
and $d$ does not simplify. 
We have 
$$a+b+c+d=(I_2-D)^T S(I_2-D) +(I_2-D)^T S D +D^TS(I_2-D)+D^T M_{22} D=\tilde{M}_{22},$$
and the expression reduces to $D^T(M_{22}-S) D=\tilde{M}_{22}-S$.  Thus, finally, we have our solution
$$
C=M_{11}^{-1} M_{12} (I_2-D), ~~D= (M_{22}-S)^{-1/2}(\tilde{M}_{22}-S)^{1/2} . $$

\noindent
The final expression is as desired, preserving singletons while rotating only the family members. 
The solution is unique if $X^T X$ is of full rank $n$.   However, in our treatment, $X$ has been row-centered, so 
no exact solution exists. To prove this by contradiction, suppose $A$ exists such that 
$A M A^T=\tilde{M }.$ When $X$ is row-centered, $M$ has rank $n-1$, and the rank of the left-hand side  cannot exceed $n-1$.
However, when matrix substitution is implemented in practice, the resulting $\tilde{M}$ typically has rank $n$, creating a contradiction.
In practice, when $X$ has been row-centered, we add a small value $\delta=0.001$ to the diagonal of $M$ before
proceeding, which provides similar results to using a Moore-Penrose generalized inverse when solving $C$ and $D$.  Either approach
results in $\frac{1}{p-1}Y^TY$ as a close approximation to $\widetilde{M}$ in simulations and for the real CF data.

\subsection{Appendix C. Simulation of genotypes}

We simulated genotype data in a manner that respected local correlation structure, which is present but typically modest in SNPs used for stratification control, and reflected population ancestry. A SNP ``block size" of 20 was chosen. An autoregressive normal model was used to simulate a set of modestly underlying correlated values, e.g. for one individual the value for the $i$th SNP is $Z_i=\rho Z_{i-1}+\epsilon$, where $\epsilon\sim N(0,1-\rho^2)$, followed by reversal of sign of $\rho$ with probability 0.5. Marginally, each $Z_i\sim N(0,1)$, and a modest $\rho=0.2$ was used within each block and $\rho=0$ at block boundaries, so that values across different blocks were uncorrelated. To convert the values to genotypes, we first generated random minor allele frequencies by drawing ``ancestral" allele frequencies from the half-triangular distribution $f(x)=2(x-a)/(a-b)^2$, where $a=0.38, b=0.50$, which corresponded closely to the observed minor allele frequency in the thinned CF dataset.
For ancestral minor allele frequency $q$, the Balding-Nichols model was used for fixation index $F_{ST}$ by drawing $K$ subpopulation allele frequencies from the beta distribution with parameters $q(1-F_{ST})/F_{ST})$, and $(1-q)(1-F_{ST})/F_{ST}$. Conversion of the latent $Z$ values to genotypes was performed by applying, for each SNP and individuals in subpopulation $k$ with allele frequency $q_k$, an inverse quantile of ranked z-values such that the lowest $z$ values were converted to genotype 0, the largest to genotype 2, and genotypes 0, 1, and 2 occurred with frequencies $(1-q_k)^2$, $2 q_k (1-q_k)$, and $q_k^2$ (i.e. Hardy-Weinberg equilibrium within each subpopulation $k$).

\bibliographystyle{agsm}
\bibliography{family}

\end{document}